# The MagE Spectrograph

J. L. Marshall[1], Scott Burles[2], Ian B. Thompson[1], Stephen A. Shectman[1], Bruce C. Bigelow[3], Gregory Burley[1], Christoph Birk[1], Jorge Estrada[1], Patricio Jones[4], Matthew Smith[2], Vince Kowal[1], Jerson Castillo[1,] Robert Storts[1], Greg Ortiz[1]

[1]Observatories of the Carnegie Institution of Washington
813 Santa Barbara Street, Pasadena, CA 91101

[2]Department of Physics and Kavli Institute for Astrophysics and Space Research,
Massachusetts Institute of Technology, 77 Massachusetts Ave, Cambridge, MA 02139

[3]UCO Lick Observatory, University of California, Santa Cruz, CA 95064

[4]Las Campanas Observatory, Casilla 601, La Serena, Chile

## ABSTRACT

The Magellan Echellette (MagE) spectrograph is a single-object optical echellette spectrograph for the Magellan Clay telescope. MagE has been designed to have high throughput in the blue; the peak throughput is 22% at 5600 Å including the telescope. The wavelength coverage includes the entire optical window (3100 Å - 1 μm). The spectral resolution for a 1" slit is R~4100. MagE is a very simple spectrograph with only four moving parts, prism cross-dispersion, and a vacuum Schmidt camera. The instrument saw first light in November 2007 and is now routinely taking science observations.

## 1. INTRODUCTION

The MagE spectrograph is a new moderate-resolution optical echellette spectrograph for the Magellan Clay Telescope at Las Campanas Observatory, Chile. MagE was built collaboratively by OCIW and MIT. The instrument has been designed to observe faint targets in the ultraviolet region (wavelengths blueward of 3600 Å), while maintaining high resolution and full wavelength coverage of the visible spectrum (3200 Å – 1 μm).

MagE is a simple, fixed format spectrograph. The optical design is very similar to the ESI spectrograph (Sheinis et al. 2002) and incorporates a reflective collimator, a medium-order ($6 < n < 20$) reflective diffraction grating in combination with two large fused-silica prisms (one single-pass, one double-pass) to provide cross dispersion. An $f/1.4$ Schmidt camera focuses the ~100 mm collimated beam onto a low-noise, back-illuminated CCD optimized for ultraviolet wavelength coverage. MagE provides resolutions of R~1000 to R~8000, using a movable slit plate containing non-overlapping slits that are 10" long and 0.5" to 5" wide.

MagE's capabilities, coupled with Magellan's 6.5m aperture, make the spectrograph especially useful for studies of the intergalactic medium, faint quasars, high-redshift galaxies, ultra-metal poor stars in our Galaxy, clusters of stars in neighboring galaxies, X-ray binaries, supernovae, gamma-ray bursts, and solar system targets of Kuiper belt objects and asteroids. It was installed on the telescope and commissioned in November 2007. MagE has already been heavily subscribed by the Magellan user community; in the 2008 calendar year MagE has been scheduled 90 nights.

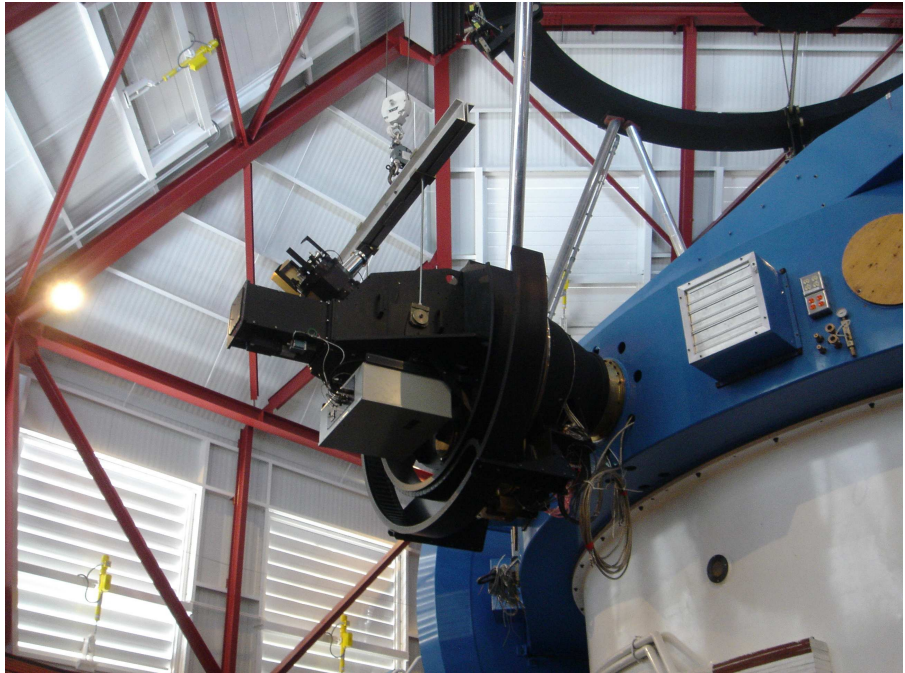

**Figure 1.** Photograph of MagE installed on the Clay folded port 2.

Figure 1 shows a photograph of MagE installed on the Magellan Clay telescope. More information about MagE is available on the MagE website at http://www.lco.cl/lco/telescopes-information/magellan/instruments-1/mage/.

## 2. OPTICAL DESIGN

The optical layout of the instrument, shown in Figure 2, is similar to the layout of the instrument ESI at Keck (Sheinis et al. 2002). The telescope focus occurs 125 mm beyond the instrument mounting surface. The collimator is an off-axis paraboloid manufactured by SORL with a focal length of 1094 mm, giving a 99 mm collimated beam diameter at the f/11.0 focal ratio of Magellan. There are two cross-dispersing prisms, both made of fused silica. The first prism, with an apex angle of 46°, is used in double pass and the second prism, with an apex angle of 40° deg, is used in single pass. The 175 gr/mm reflecting grating, which is used in the quasi-Littrow configuration, has a 102x128 mm ruled area and a blaze angle of 32.3° (6.2 μm first order Littrow blaze). The grating provides a resolution of R=4100 for a 1 arcsec slit on Magellan.

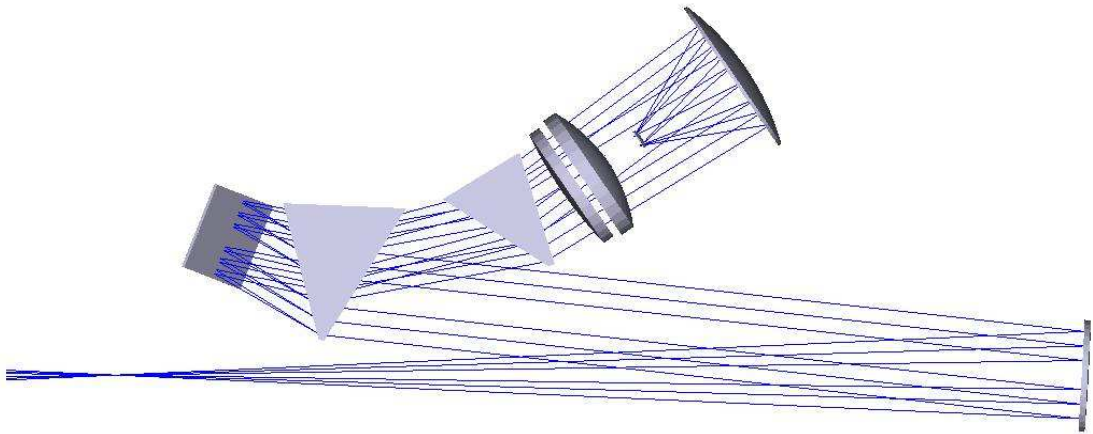

**Figure 2.** MagE optical design.

The rotation angles of the two prisms have been adjusted to control the anamorphic beam distortion. The grating is used in orders 6 through 20 and the prisms provide adequate cross-dispersion to separate all orders when used with a 10 arcsec slit. Figure 3 shows a solar spectrum taken with MagE with the echelle orders and central wavelengths of each order marked.

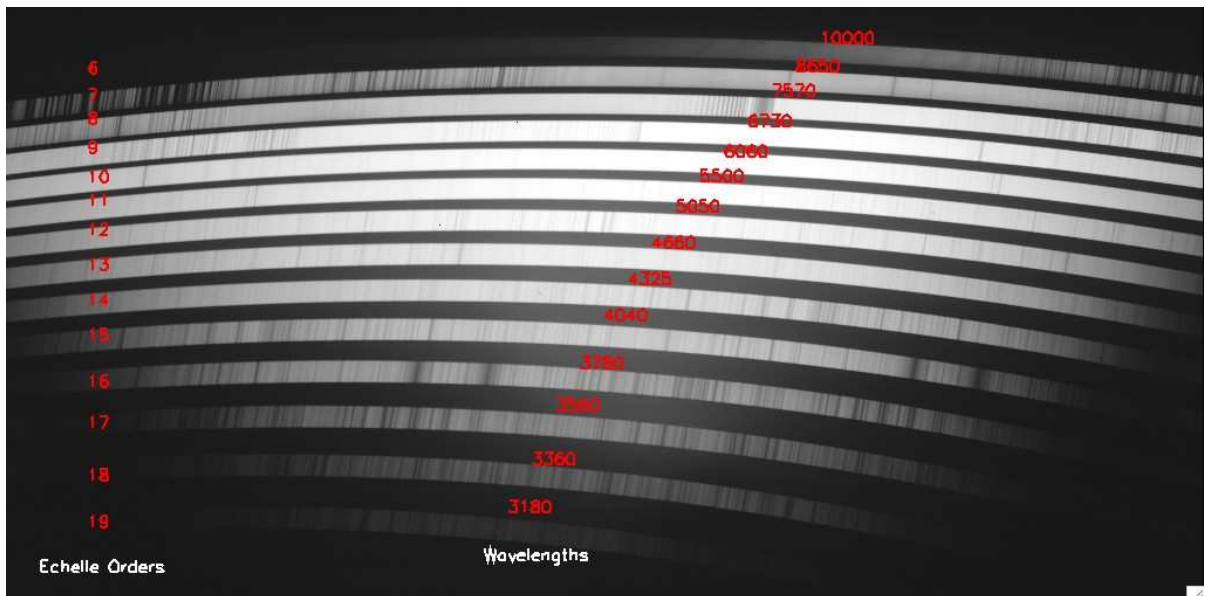

**Figure 3.** Solar spectrum taken with MagE. The central wavelength of each order and the order number are indicated.

The 2048x1024 CCD detector, with 13.5 μm pixels, is placed at the prime focus of a vacuum Schmidt camera with a focal length of 138 mm. The angular field of view of the camera is 12.7°; the scale at the detector is 3 pixels per arcsec on the sky. The first element of the camera is an aspheric $CaF_2$ meniscus lens manufactured by Janos Technology. The aspheric surface was produced by single-point diamond turning without post-polishing and has an rms surface ripple of ~2 nm. The second element of the camera is a spherical fused silica meniscus which is the vacuum window of the camera. The spherical mirror is 200 mm in diameter and has a radius of curvature of 293 mm. A fused silica field flattener with spherical surfaces is placed 2 mm in front of the detector. The field flattener is trimmed to a minimum rectangular shape (33.6x18.8 mm) to minimize obscuration of the incoming beam.

The CCD detector and field flattener are mounted at a fixed position on a G-10 frame attached directly to the fused silica meniscus, which has a small invar insert epoxied into a shallow recess at the center of the inside surface. The critical mechanical adjustments are the lateral position of the aspheric $CaF_2$ meniscus, and the tip, tilt and focus of the camera mirror, which can be adjusted from outside the camera assembly via vacuum feed-throughs. The positions of the corrector and the mirror are adjusted iteratively in order to achieve the best possible images across the full area of the CCD. The measured image diameter (using a 0.35 arcsec square entrance aperture) is less than 2.0 pixels FWHM everywhere in the field; the average image diameter is 1.8 pixels.

The collimator and camera mirrors were both coated at LLNL using a high-performance enhanced-UV reflecting coating with an average reflectivity of ~94% between 3200A and 1 μm (Thomas and Wolfe 2000). The antireflection coatings on the cross-dispersing prisms are made of Sol-gel over MgF, provided by Cleveland Crystals. The fused silica meniscus and the field flattener were coated with a broad-band hard dielectric AR coating by ZC&R. In order to reduce scattered light, there are 3 baffle rings placed inside the camera between the fused silica meniscus and the mirror, in addition to the two aperture stops. All were coated with Lord Aeroglaze Z306 flat black polyurethane, applied by Barry Avenue Plating.

# 3. INSTRUMENT DESIGN

The MagE mechanical and opto-mechanical design were guided by principles intended to achieve the following: sub-system modularity, minimize mass, minimize image motion (flexure), maintain the lowest practical part counts (e.g. use of part symmetry), ease of fabrication, ease of assembly, and ease of serviceability. Figure 4 shows a solid model image of the instrument.

## 3.1. Instrument Structure

The instrument mainframe is a bolted aluminum plate structure. The mainframe provides external interfaces to the Magellan instrument hoist, the Magellan folded-port instrument rotator, the collimator assembly, the Schmidt camera, the calibration system, the slit viewing camera, and the instrument electronics enclosure. Internal interfaces are provided for mounting the slit stage, internal baffles, grating mount, and prism mounts. Dove-tailed joints in the instrument structure prevent light leaks through the plate intersections. Internal baffles isolate the slit-viewing and calibration optical paths from the spectrograph volume. The total weight of the bare instrument structure is 300 lbs.

The instrument structure includes an external electronics enclosure, described in more detail in Section 4.3, and a utility wrap to supply power, plumbing, cryocooler lines, ethernet, and optical fiber to the instrument. These utilities are fed from the telescope to the instrument by a single layer IGUS energy chain. The chain allows approximately 540 degrees of rotation of the instrument and features quick-disconnect fittings at each end to facilitate removal and installation of the instrument on the telescope.

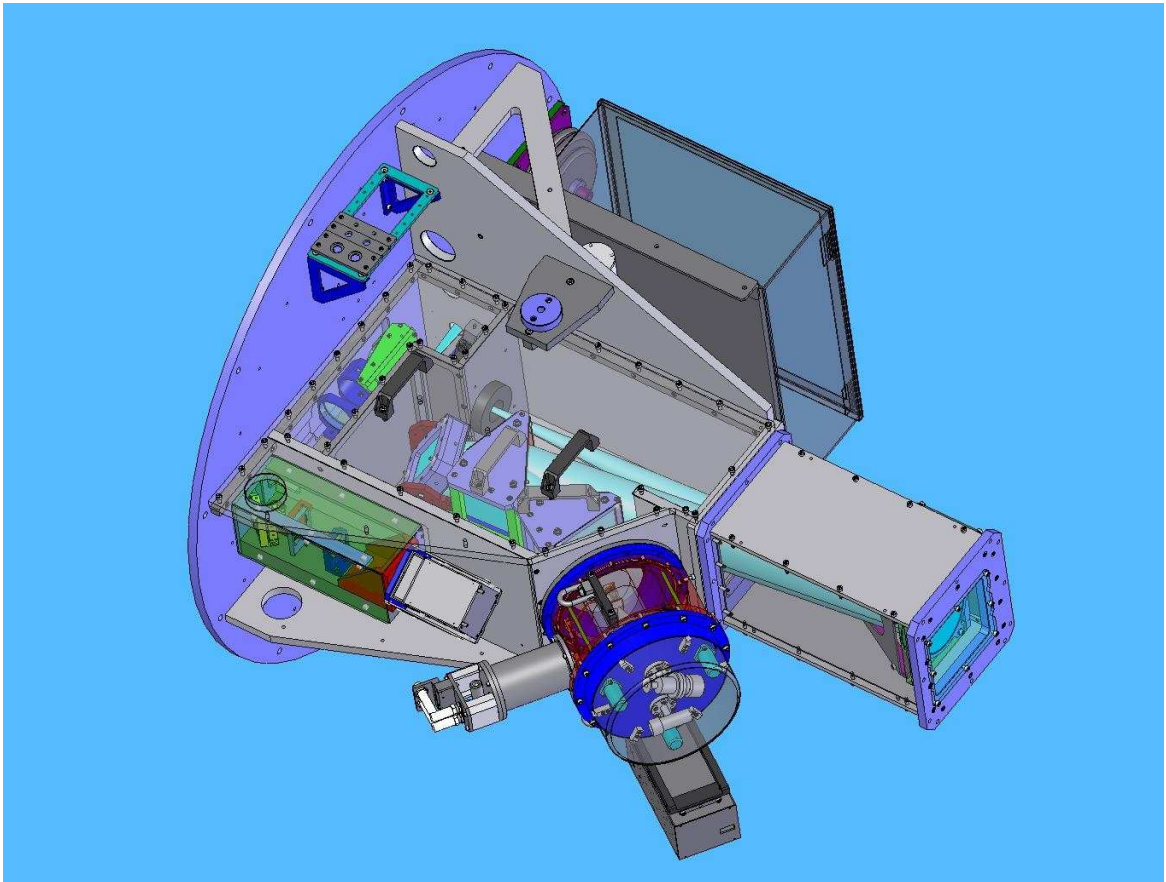

**Figure 4.** Solid model image of instrument

## 3.2. Collimator Assembly

The MagE collimator assembly consists of an off-axis paraboloid mirror kinematically mounted in a thermally compensated aluminum cell that is carried on a linear bearing carriage. Collimator focus may be adjusted by the observer through the instrument control software. The collimator mirror carriage is spring-loaded against the actuator through hardened steel contacts and preloaded such that the focus carriage remains in contact with the actuator for all orientations of the instrument. The collimator assembly is mounted via an interface flange to the instrument structure.

## 3.3. Grating and Prism Mounts

The echelle grating is mounted in a simple kinematic cell, in which nylon-padded defining points are pre-loaded by nylon-tipped spring plungers. Fine tip/tilt adjustment of the grating is accomplished with precision shims between the cell and active surface of the grating. Gross alignment tilts are set by vertical and horizontal pivots built into the cell and its mountings.

The two cross-dispersing prisms are mounted (identically) in kinematic cells, again with nylon-padded defining points and nylon-tipped spring plungers. Three of the defining and preloading points act on the two triangular surfaces of the prisms. Two fixed defining points act on the first optical surface, and a second pair of whiffle-tree pivoting contacts act on the second. Effectively, these four contacts provide the remaining three constraints, and are preloaded by spring plungers acting on the surface opposite the prism vertex.

## 3.4. Vacuum Schmidt Camera

The MagE science camera is an unusual design, in which the second of the two Schmidt corrector lenses also acts as the window to the science detector vacuum vessel. The detector is mounted to the backside of the second corrector via an Invar puck which is epoxy-bonded to a recess in the center of the lens. A G-10 fiberglass hexapod truss thermally isolates the cooled CCD detector from the ambient-temperature vacuum lens. Figure 5 shows a photograph of this assembly. Athermalizing Invar rods acting between the detector and the Schmidt primary mirror maintain camera focus over all operating temperatures. The Schmidt corrector lens is potted into an aluminum cell with RTV silicone adhesive, which allows X-Y alignment with respect to the optical axis by way of nylon push screws. When alignment is complete, the front corrector is locked in place by a second set of retaining screws.

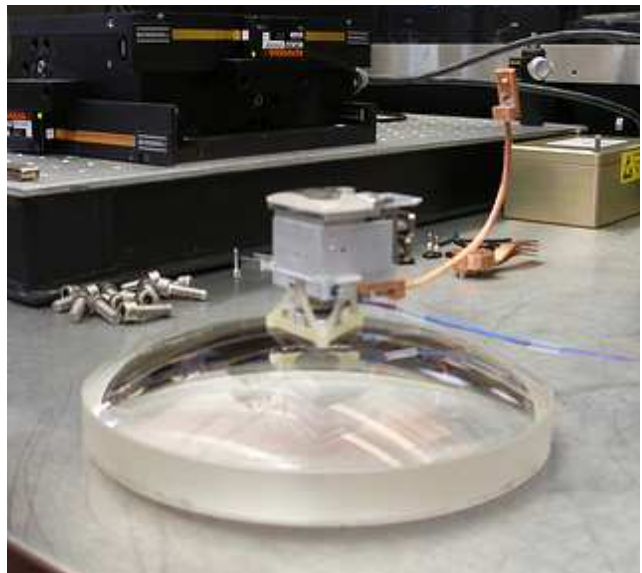

**Figure 5.** Photograph of the CCD assembly mounted to the Schmidt corrector lens.

The athermal design of the Schmidt camera system obviates the need for operator focus of the camera. However, the camera optics are extremely sensitive to misalignment of the detector and field flattener with respect to the aspheric corrector and the spherical Schmidt primary mirror. For this reason, ferro-fluidic feed-throughs allow for tip/tilt/piston alignment of the mirror under vacuum conditions. The mirror cell is attached to the camera barrel via a laser-cut diaphragm flexure, which constrains translation and rotation in the plane of the mirror while allowing tip/tilt/and piston to be defined by the Invar metering rods.

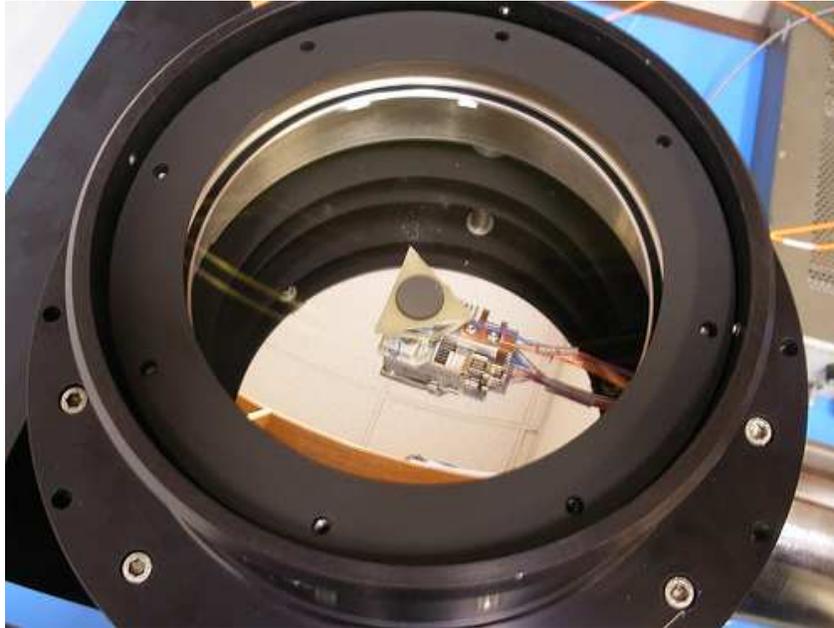

**Figure 6.** Photograph of Schmidt camera looking through the dewar window, showing the back of the mounting puck for the CCD assembly, the CCD assembly and associated cables, the black baffle rings, and the primary mirror.

The camera body is a vacuum vessel composed of a tube sealed at the front with one of the Schmidt corrector lenses and at the back with a plate that supports the camera primary mirror. The detector assembly is supported by the Schmidt corrector lens, and the lens and all other entrance ports to the vacuum are sealed with Viton O-rings. The vessel is black anodized outside and polished on the inside to minimize radiative heat exchange between the cold detector and the ambient temperature vessel. An acceptable vacuum level is maintained with an ion pump operating at all times. A photograph of the camera is shown in Figure 6.

### 3.5. Calibration System

The calibration system includes an optical system which reproduces the *f*-ratio and focal position of the telescope. An automated New Focus flipper mirror switches the input beam between the telescope and the calibration system. Lenses and mirrors in the optical system are mounted in commercial opto-mechanical mounts. Calibration sources (hollow cathode and flash lamps) illuminate an integrating sphere, which in turn outputs through a pin-hole to the optical system. The calibration system mounts to the outside of the instrument structure, to minimize stray light and ease servicing of the calibration sources.

The calibration system contains a Thorium-Argon hollow cathode tube and a flat field flash lamp. The hollow cathode tube is supplied for wavelength calibration and provides suitable lines over the entire wavelength range. Typical wavelength solutions give an RMS of 0.06 A for 500 arc lines. The flat field calibration source is a Xe-flash lamp. The Xe-flash lamp provides pulses of light at a well-calibrated frequency. One drawback of this lamp is that it contains broad emission lines that must be filtered out before applying the flat field. We have found that it is most effective to use the Xe-flash lamp to calibrate the blue orders of the spectrum, and to use the dome flat field system with incandescent illumination available at the telescope to calibrate the red orders.

### 3.6. Slit Mechanism and Slit Viewing System

The MagE slit plate is a carefully constructed component. The slits are etched into a 0.003 inch thick NickelCobalt plate using photolithography and liquid etching. The slit plate is then mounted on a stage that is controlled by a stepper motor. The slit plate is tilted with respect to the science beam in order to redirect light to the slit viewing camera.

The observer may select from a range of available slit widths: 0.5, 0.7, 0.85, 1.0, 1.2, 1.5, 2.0, and 5.0 arcsec. There is also a position with three 0.35 arcsec pinholes used for focusing the spectrograph. All slits are 10 arcsec long with a plate scale of 0.3 arcsec per pixel on the detector.

The slit viewing camera allows the observer to see a ~1 arcminute image of the region near the slit. The slit viewing optics include both commercial and custom lenses, each of which are bonded with RTV silicone adhesive into aluminum cells. These optics produce excellent image quality at the detector. The optical path includes a filter which may be exchanged manually while the instrument is mounted to the telescope. A standard Magellan glycol-cooled CCD guide camera provides the detector system for the slit viewer.

## 4. DETECTORS, SOFTWARE, & INSTRUMENT CONTROL

### 4.1. CCD System

The MagE CCD is an E2V 42-20 back-illuminated device, which has a 2048x1024 format with 13.5 μm pixels. It is coated with a broadband anti-reflective coating. The MagE CCD shows fringing in the red region of the spectrum, starting at about 7000Å, with peak fringing amplitudes reaching about 10 per cent.

The detector control electronics are very similar to those of other Las Campanas instruments. The CCD controller is a compact, two-channel digital signal processor based design, and is read out through one channel only. More information about the CCD control electronics may be found at http://www.ociw.edu/instrumentation/ccd/base/base.html. The CCD may be read out at three speeds with different gain and readnoise characteristics. These are summarized in Table 1.

**Table 1.** MagE readout characteristics

| Read speed | Read time (s) | Gain (e-/DN) | RMS noise (e-) |
|:---:|:---:|:---:|:---:|
| Slow | 33 | 1.02 | 2.9 |
| Fast | 21 | 0.82 | 3.1 |
| Turbo | 14 | 1.53 | 4.2 |

The Mage CCD is cooled by a single Cryotiger closed-cycle cooler to an operating temperature of −110° C. The cooling path between the Cryotiger and the detector includes an activated charcoal getter and a magnetic coupling, which allows simple disassembly of the cooler head from the CCD.

### 4.2. Control Software

The MagE control software operates on the observer computers in the telescope control room. The software and User GUI are very similar to those of many other Magellan instruments. One special feature of the MagE software is a QuickLook tool that automatically displays each image obtained, and then allows the user to inspect particular features in the spectrum by displaying a wavelength associated with the cursor position in the displayed image. This allows the observer to easily inspect individual features in real time without using additional software.

### 4.3. Control Electronics

The instrument control electronics are housed in a small Hoffman enclosure that is mounted on the outside of the instrument structure. MagE has very few automated functions and thus requires very little volume, mass, or cooling for the electronics. The electronics enclosure carries power supplies for the AC synchronous motors that drive the slit mechanism and collimator focus, as well as power supplies for the camera electronics and calibration sources.

The instrument electronics and mechanisms are controlled by a programmable logic controller (PLC). An AutomationDirect DL205 (260 CPU) is commanded either from the instrument's GUI or its local control panel. The local panel allows the instrument scientists to control every mechanism of the instrument, with the exception of the CCD controller. The PLC has discrete and analog input modules, discrete output modules, and an Ethernet communications module which enables it to communicate through the network, either from the observer's computer or via the PLC's programming software, which allows it to be monitored and debugged on line. The PLC controls all of the instrument mechanisms: slit plate, focus control, shutter, calibration lamps, flipper mirror, and flash lamp. It also monitors temperature inside the instrument and controls the ion pump.

## 5. PERFORMANCE

### 5.1. Throughput

The throughput of MagE was measured on the night of 23/24 November 2007. A total of 10 spectrophotometric standards were observed over an airmass range of 1.04 to 2.63. Table 2 gives the results of these measurements. In the table the order number is followed by the central wavelength of the order and the dispersion at that wavelength. This is followed by the zero point magnitude at that wavelength for one air mass, the extra-order flux in magnitudes (to be added to the zero point magnitude), and the measured extinction. Finally the table lists the overall efficiency in per cent for the telescope plus instrument, and for the instrument alone (assuming three telescope mirror reflections of 0.85 per cent at all wavelengths).

**Table 2.** Throughput measurements

| Order | Lambda (Å) | Dlambda (Å) | Zeropoint (1 ct/s/Å) | Extra Flux | Extinction | Efficiency | Efficiency (instrument) |
|---|---|---|---|---|---|---|---|
| 20 | 3125 | 0.231 | 17.504 | 0.191 | 1.523 | 0.091 | 0.149 |
| 19 | 3260 | 0.244 | 18.434 | 0.046 | 0.952 | 0.116 | 0.189 |
| 18 | 3440 | 0.258 | 18.791 | 0.036 | 0.587 | 0.120 | 0.196 |
| 17 | 3650 | 0.274 | 19.042 | 0.042 | 0.464 | 0.144 | 0.235 |
| 16 | 3860 | 0.292 | 19.212 | 0.025 | 0.370 | 0.161 | 0.263 |
| 15 | 4130 | 0.311 | 19.438 | 0.026 | 0.291 | 0.198 | 0.322 |
| 14 | 4400 | 0.335 | 19.461 | 0.024 | 0.215 | 0.200 | 0.326 |
| 13 | 4750 | 0.360 | 19.491 | 0.007 | 0.176 | 0.211 | 0.344 |
| 12 | 5140 | 0.390 | 19.503 | 0.000 | 0.114 | 0.217 | 0.353 |
| 11 | 5590 | 0.426 | 19.415 | 0.000 | 0.117 | 0.218 | 0.355 |
| 10 | 6130 | 0.469 | 19.316 | 0.000 | 0.085 | 0.212 | 0.345 |
| 9 | 6800 | 0.521 | 19.125 | 0.000 | 0.028 | 0.187 | 0.305 |
| 8 | 7520 | 0.590 | 18.682 | 0.000 | 0.004 | 0.135 | 0.219 |
| 7 | 8610 | 0.674 | 18.057 | 0.000 | 0.149 | 0.099 | 0.161 |
| 6 | 9700 | 0.788 | 16.300 | 0.000 | 0.027 | 0.020 | 0.032 |

## 5.2. Flexure

We have measured the amount of image motion caused by instrumental flexure at a range of instrument rotations and telescope elevations. For these tests, the slit was positioned to place the three pinholes in the beam. We measured instrumental flexure in two axes: by rotating the instrument about the axis of the instrument rotator and by moving the telescope to different elevations. We measured image motions of about 0.65 pixels when the instrument was rotated from 0 to ±175 degrees, and by at most 0.8 pixels when the telescope was moved from 90 degrees elevation (zenith-pointing) to 20 degrees (nearly horizon-pointing). Figure 7 shows these results.

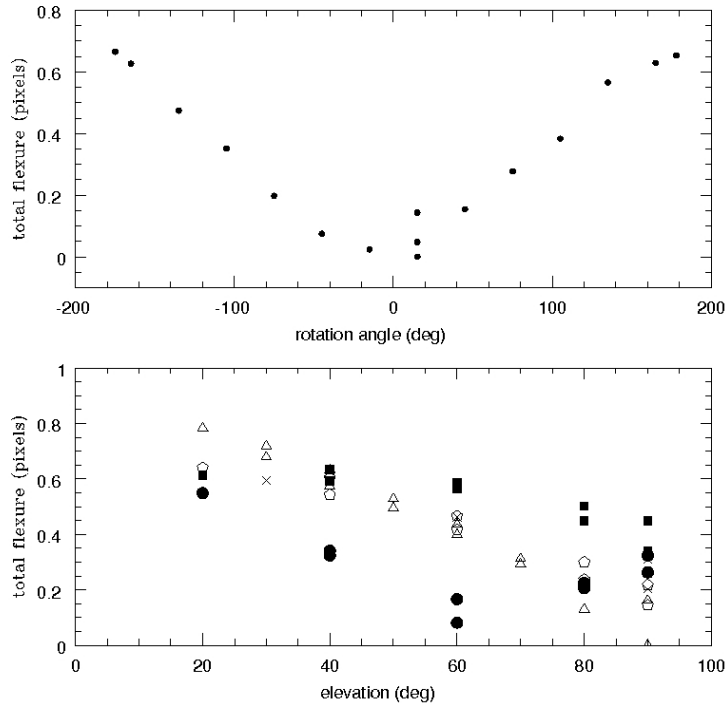

**Figure 7.** MagE flexure measurements. The instrument flexes both as the instrument rotates around the axis of the rotator and as the telescope changes elevation, but in all cases by less than one pixel (13.5 μm or 0.3 arcsec). In the lower plot, the symbols correspond to the same test repeated at different instrument rotations.

Flexure of less than 1 pixel over these rotations should have a negligible impact on science data quality, unless the target is near the zenith and therefore the instrument is rotating at a high rate. For example, while the slit is aligned with the parallactic angle and the telescope tracks over 30 degrees in elevation the maximum flexure is 0.1 pixel.

## 6. SUMMARY

MagE is a new, moderate-resolution optical echellette spectrograph that is optimized for faint ultraviolet observations. The spectrograph has been fully commissioned and deployed at the Magellan Clay telescope, and is currently available for use by the Magellan user community.

## ACKNOWLEDGEMENTS

This project was funded in part by NSF grant AST-0215893.